\documentclass[english,pra,twocolumn,showpacs]{revtex4}
\usepackage{amsmath}
\usepackage{graphicx}
\usepackage{amssymb}

\begin{document}
\title{Noise Suppression for Micromechanical Resonator via Intrinsic Dynamic
Feedback}

\author{H. Ian}
\email{hian@itp.ac.cn}
\author{Z. R. Gong}
\author{C. P. Sun}
\email{suncp@itp.ac.cn}
\homepage{http://www.itp.ac.cn/suncp}
\affiliation{Institute of Theoretical Physics, The Chinese Academy of Sciences,
Beijing, 100080, China}
\date{\today}

\begin{abstract}
We study a dynamic mechanism to passively suppress the thermal noise
of a micromechanical resonator through an intrinsic self-feedback
that is genuinely non-Markovian. We use two coupled resonators, one
as the target resonator and the other as an ancillary resonator, to
illustrate the mechanism and its noise reduction effect. The intrinsic
feedback is realized through the dynamics of coupling between the
two resonators: the motions of the target resonator and the ancillary
resonator mutually influence each other in a cyclic fashion. Specifically,
the states that the target resonator has attained earlier will affect
the state it attains later due to the presence of the ancillary resonator.
We show that the feedback mechanism will bring forth the effect of
noise suppression in the spectrum of displacement, but not in the
spectrum of momentum. 
\end{abstract}


\maketitle

\section{INTRODUCTION}

Recently, interests have been generated on cooling techniques for
mechanical systems at nano- and micron-scales \cite{karrai06}. Among
them, the typically employed is the feedback cooling technique where
an external feedback circuit is responsible for detecting the motion
of the target and feeding a counteracting force against this motion;
through a general decrease of magnitude in the density noise spectrum,
it was shown that the feedback can effectively reduce the fluctuation
of the target and provide a cooling mechanism \cite{hopkins03,durso03}.
Some experiments based on the models that contain the feedback loops
have been carried out in the past few years. A few are directed towards
the cooling of micron- to nanometer-size mechanical resonators, aiming
to reach a macroscopic quantum mechanical ground state and serving
as a powerful manifestation of quantum mechanical effects \cite{cohadon99,naik06,lahaye04,pzhang05}.
Other experiments succeed in slowing down the motion of micron-size
mirrors through the radiation pressure of an optical field in a Fabry-P\'{e}rot
(FP) cavity, aiming to reach a noise level and equivalently an effective
temperature that are pertinent to the employment of high-precision
detection of gravity waves \cite{metzger04,arcizet06,gigan06,kleckner06}.

The forementioned implementations of feedback cooling through reduction
of noise fluctuations invariably rely on an electrical circuitry external
to the target system to be cooled. The controller here is usually
fixed and attracts or repels the resonator through either electrostatic
Coulomb force or Lorentzian force. If such an external detection-control
unit could be eliminated in favor of a mechanism with self-detection
of and self-adjustment to the target's thermal motion, we call the
mechanism {}``self-cooling'' \cite{bhatt07,ydwang07}. Devices implementing
this self-cooling use less components and are free from the reliance
on an external circuitry and hence prone to less noise sources.

In one case \cite{bhatt07}, an augmented cavity along with an extra
optical cavity field is established on the other side of the mirror,
in addition to the regular FP cavity, so as to counteract the radiation
pressure from the original cavity field. This extra field cushions
the motion of the pressure mirror and plays the role of feedback.
In another case \cite{fxie07,ydwang07}, a set of Josephson junctions
behaving as a qubit, serially connected to a mechanical resonating
beam, serves delayed supercurrent into the circuit according to the
magnetic flux through the circuit loop. The magnetic flux is controlled
by the vibrating motion of the beam, which in turn is controlled by
the magnetic field generated by the current feed. Such a mutual dependence
furnishes a self-feedback mechanism. It should be pointed out that
both of the self-feedback setups require delayed feedback, which assumed
\textit{a priori} a non-Markovian approximation that explicitly depends
on the history of the target's motion. In these phenomenological treatments,
the cooling target either couples itself to a static controller and
makes itself prone to the noise stemmed from the feedback, or couples
to a mechanically static detection construct and receives manually
delayed feedback.

Hereby, we present a dynamic model based on an intrinsic mechanism
with non-Markovian feedback, which is obviously free from an external
feedback loop and does not rely on a presupposition of historical
dependence. This mechanism is illustrated by a simple mechanical system
in which the target is modeled by a harmonic oscillator and attached
to a dynamic controller, which is a relatively heavier resonator,
through a spring. The target is controlled by an intrinsic feedback
through the dynamics of coupling: earlier positions and velocities
of the target affects the motion of the controller and this influence
is subsequently fed back to the target. Consequently, the accumulation
of earlier states of the target will affect the state of itself later.
With proper parameter setup, the target essentially experiences a
resistance and decelerates its motion; its displacement variance is
shrunk, noise suppressed and effective temperature cooled down. The
lack of a specific detection device for the motion of the target resonator
and an external feedback circuit characterizes the intrinsic nature
of the mechanical feedback. Our numerical analysis shows the existence
of a noise suppression capability of our scheme, e.g. the variance
of displacement can be reduced to $0.04\times10^{-21}m^{2}$, and
therefrom a cooling capability under a practical setting accessible
in current experiments. We note that the scheme is theoretically illustrative
through its simple model setup yet widely applicable because the general
oscillator systems can be extended to quantum bosonic systems and
other cases. In fact, a similar model and mechanism has been proposed
to actively cool down the torsional vibration of a nanomechanical
resonator through spin-orbit interactions \cite{nzhao07}.

The model will be explained in Sec.\ref{sec:model} and its delay
function then derived \textit{a posteriori} to examine its non-Markovian
dependence. The complete solution of the system dynamics is given
in Sec.\ref{sec:exact_soln}, with which we will derive the density
noise spectrum and calculate the theoretical noise suppression rate.
The associated numerical results will be presented in Sec.\ref{sec:num_analysis},
given various parameter setups. The analysis is extended to the domain
of momentum noise in Sec.\ref{sec:momentum-noise}.

%%%%%%%%%%%%% figure of the model %%%%%%%%%%%%%%%
%
\begin{figure}
\begin{centering}
\includegraphics[width=3in]{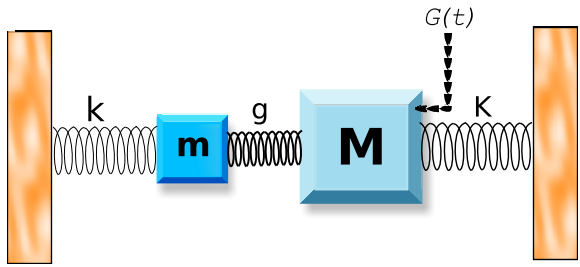} 
\par\end{centering}

\caption{(Color online) The diagrammatic figure shows the arrangement of the
three springs and the two masses. From left to right, they are: the
spring of constant $k$, the target mass $m$, the spring of constant
$g$, the ancillary mass $M$, and the spring of constant $K$. $G(t)$
is the harmonic driving force.}

\label{fig:model} 
\end{figure}

%%%%%%%%%%%%% end of figure %%%%%%%%%%%%%%%%%%%%%

\section{\label{sec:model}INTRINSIC FEEDBACK BY COUPLING DYNAMICS}

\subsection{The Model}

Our model setup (see Fig.\ref{fig:model}) comprises two masses and
three springs. The two masses are denoted by $m$ and $M$, respectively.
The mass $m$ is the target and typically lighter whereas the mass
$M$ serves as an ancillary controller and is relatively heavier.
The three springs are denoted by their Hooke's constants $k$, $g$
and $K$, respectively. The spring of constant $k$ attaches the lighter
mass $m$ to the fixed wall on the left and the spring of constant
$K$ attaches the heavier mass $M$ to the fixed wall on the right.
The spring of constant $g$ strings the two masses together. Such
a setup, intuitively, grants the heavier mass $M$ the function of
a suspension system and a medium for the feedback. The symbol $G(t)$
represents an external driving force which is necessary for the discussion
of cooling but can be deemed zero for the present.

The connected springs will give rise to mechanical vibrations of the
masses. We let $\bar{\omega}=\sqrt{(k+g)/m}$ denote the effective
mechanical resonance frequency for the mass $m$, assuming the other
mass $M$ is fixed, and $\bar{\Omega}=\sqrt{(K+g)/M}$ the equivalent
for the mass $M$, assuming the mass $m$ is fixed. Besides these
mechanical vibrations, we assume each of the masses experience a frictional
damping and we let $\gamma$ denote the damping coefficient for the
mass $m$ and $\Gamma$ that for mass $M$.

Then according to the setup above, the coordinates of the two masses
obey a coupled system of classical Langevin equations \begin{eqnarray}
\ddot{x}+\gamma\dot{x}+\bar{\omega}^{2}x-\frac{g}{m}Q & = & f\label{eq:langevin_x}\\
\ddot{Q}+\Gamma\dot{Q}+\bar{\Omega}^{2}Q-\frac{g}{M}x & = & F\label{eq:langevin_Q}\end{eqnarray}
 where $x$ is the coordinate of the mass $m$ and $Q$ that of the
mass $M$. $f$ and $F$ on the right hand side of the equations denote
the random thermal noise generated by the mass $m$ and $M$, respectively,
due to their frictional damping.

The frictional damping terms as dissipation and the random noise terms
as fluctuations constitute the total Brownian thermal force in the
classical Langevin formalism. This thermal force induces an thermal
environment, the effect of which is divided among the two terms $f(t)$
and $F(t)$ according to the fluctuation-dissipation relations \cite{kubo85}
\begin{eqnarray}
\left\langle f(t)f(t^{\prime})\right\rangle  & = & 2k_{B}T\frac{\gamma}{m}\delta(t-t^{\prime}),\label{eq:FD-relation}\\
\left\langle F(t)F(t^{\prime})\right\rangle  & = & 2k_{B}T\frac{\Gamma}{M}\delta(t-t^{\prime}).\label{eq:FD-relation_F}\end{eqnarray}
 For now we assume the two masses are independently interacting with
two thermal environments. That is, there are no correlations between
the fluctuations of the two masses

Actually, we can realize from Eq.(\ref{eq:langevin_x}) and Eq.(\ref{eq:FD-relation})
that the motion of the mass $m$ is resisted by a frictional force
$\gamma$, which in turn is transduced into thermal energy and heats
up itself. This process, however, is mediated by the mass $M$ that
stands between $m$ and the fixed wall through the term $(g/M)Q$.
If there were not the mass $M$, the kinetic oscillation of $m$ would
be instantly reacted by the surrounding springs of Hooke's constant
$k$ and $K$. With the presence of the mass $M$ and the extra spring
of constant $g$, the oscillation of the mass $m$ will first squeeze
the spring of constant $g$, and then the squeezed spring will release
and push the mass $M$ to the right. It follows that the spring of
constant $K$ will be squeezed successively. The mediating mass $M$
breaks the original single spring into two and permits these two springs
to stay in different states, squeezed and released, and hence essentially
delays the reaction of the springs. This cascaded process is then
reflected by the wall and executed in a reversed order to the left;
the delayed reaction of the springs of constant $g$ and $K$ acts
back onto the oscillating mass $m$. The delayed reaction depends
on the oscillating velocity of the target mass $m$ and can thus be
considered a feedback onto itself.

\begin{figure}
\begin{centering}
\includegraphics[width=8cm]{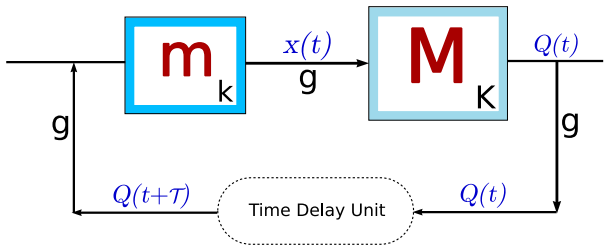} 
\par\end{centering}

\caption{Schematic illustration of the feedback mechanism shows the controller
$M$ receives signal $x(t)$ and outputs signal $Q(t)$ according
to the motion of the target $m$.}

\label{fig:feedback_flow} 
\end{figure}

The entire process can be regarded as a feedback loop from the view
of control theory through the flow diagram shown in Fig.\ref{fig:feedback_flow}.
The target resonator $m$ is the system to be controlled and the ancillary
resonator $M$ becomes the controller which detects the signal $x(t)$
as its input and feeds the signal $Q(t)$ as its output. The cycle
time of the loop corresponds to the delay of the controller-to-system
reaction. The dependence of this delayed reaction on the tunable parameters,
mainly those spring constants, allows us to control this self-feedback
precisely to counteract the oscillating motion of the target mass.

%%%%%%%%%%%%%%%%%%% figure of Feedback%%%%%%%%%%%%%%%%%%%
%

%%%%%%%%%%%%%%%%% end of figure %%%%%%%%%%%%%%%%%%%%%%%%%

The weakened motion is converted to a reduction of effective temperature
of the target mass through an equivalent relation between the autocorrelation
of the target's displacement and temperature, derived from the fluctuation-dissipation
relation Eq.(\ref{eq:FD-relation}). Under normal circumstances, that
is, when the coupling mass $M$ were not present, this equivalent
relation can be written as \begin{equation}
\left\langle x(t)x(t^{\prime})\right\rangle =2\pi k_{B}\gamma T\xi_{0}(t-t^{\prime})\end{equation}
 where $\gamma$ is defined as in Eq.(\ref{eq:langevin_x}) and $\xi_{0}$
is a function to be given explicitly. Experimentally, the statistical
variance of the coordinate is measured and the above relation is used
to compute an effective temperature.

Increasing the damping coefficient $\gamma$ will certainly increase
the enveloping rate of the oscillating motion of $x(t)$. However,
this reduction of motion cannot lead to an equivalent noise suppression
effect upon the target mass $m$ because of the constraint imposed
by the friction-induced fluctuation relation described above. In fact,
according to this relation, increasing damping rate leads to more
thermal dissipation of the system. That is to say, simply enlarging
frictional force merely results in a heating effect upon the target
system. Our aim, therefore, is to reduce the temperature of the target
mass by reducing its coordinate variance through the feedback that
does not simultaneously increase the damping coefficient. After we
derive the explicit feedback response below, we will show the feedback
is actually shrinking the variance in Sec.\ref{sec:exact_soln}.

\subsection{The Feedback}

Generally, a feedback external to an oscillating system has the effect
of adding an extra driving force term on the right hand side of the
equation of motion, i.e. \begin{equation}
\ddot{y}+\gamma\dot{y}+\bar{\omega}^{2}y=f+\mathcal{F}_{\mathrm{fb}}\end{equation}
 where we have let $y$ denote the coordinate of a general dynamic
system, $\mathcal{F}_{\mathrm{fb}}$ the feedback force, and $\gamma$,
$\bar{\omega}$, $f$ terms of similar meanings to those defined in
Eq.(\ref{eq:langevin_x}). If the target were to be cooled, the motion
of the system should be slowed down. In other words, behaving as a
function of the target's velocity, the force $\mathcal{F}_{\mathrm{fb}}$
should have an equivalent effect of increasing the damping coefficient,
but do not increase the fluctuations. Besides, the feedback force
should be dependent on the target's position and the noise source.

Summing up these requirements, we expect $\mathcal{F}_{\mathrm{fb}}$
to be a function of $y(t)$, $\dot{y}(t)$ and $f$. To overcome the
generic fluctuation-dissipation relations and hence accomplish an
efficient noise suppression, we shall use a non-Markovian type feedback:
$\mathcal{F}_{\mathrm{fb}}$ not only depends on the current value
of the velocity and the position of the target, but also their historical
values at past times. If we represent the historical dependence by
a time-derivative $\mathrm{d}K(y,\dot{y},\tau)/\mathrm{d}\tau$ and
weigh the contribution of the histories by a delay function $h(t-\tau)$,
where $t$ stands for the current time and $\tau$ for the time in
the past, the feedback force can be written as an integral with respect
to $\tau$, \begin{equation}
\mathcal{F}_{\mathrm{fb}}=\int_{-\infty}^{t}\mathrm{d}\tau\frac{\mathrm{d}K(y,\dot{y},\tau)}{\mathrm{d}\tau}h(t-\tau).\label{eq:feedback_form}\end{equation}

The above formula shows the mathematical character of a general feedback
force. Inversely, any function that can express the same character
should be considered a feedback force. Therefore, we can verify the
dynamic response of the coupling mass $M$ in our model as an effective
feedback mechanism by finding the corresponding specific expression
for $K(y,\dot{y},t)$ and $h(t-\tau)$. To do so, we solve Eq.(\ref{eq:langevin_Q})
by using Fourier transforms and integration by parts (the detailed
derivation is given in Appendix \ref{sec:derv_delay}) \begin{eqnarray}
Q(t) & = & \frac{1}{2\pi}\int_{-\infty}^{\infty}\mathrm{d}\tau\mathrm{d}\omega\frac{\frac{g}{M}x(\tau)+F(\tau)}{(\bar{\Omega}^{2}-\omega^{2})+i\Gamma\omega}e^{-i\omega(\tau-t)}\\
 & = & \frac{1}{\bar{\Omega}^{2}}\phi(x,t)+\frac{2}{\sqrt{4\bar{\Omega}^{2}-\Gamma^{2}}}\times\label{eq:soln_Q}\\
 &  & \int_{-\infty}^{t}\mathrm{d}\tau\frac{\mathrm{d}\phi(x,\tau)}{\mathrm{d}\tau}h(t-\tau)\nonumber \end{eqnarray}
 where \begin{equation}
\phi(x,t)=\frac{g}{M}x(t)+F(t)\label{eq:phi(x,t)}\end{equation}
 denotes the inhomogeneous part of the equation, i.e. the external
driving force to $M$. In the solution, \begin{eqnarray}
h(\xi) & = & -\frac{1}{2\bar{\Omega}^{2}}\exp\left[-\frac{1}{2}\Gamma(\xi)\right]\left\{ \sqrt{4\bar{\Omega}^{2}-\Gamma^{2}}\times\right.\label{eq:delay}\\
 &  & \left.\cos\left[\frac{\xi}{2}\sqrt{4\bar{\Omega}^{2}-\Gamma^{2}}\right]+\Gamma\sin\left[\frac{\xi}{2}\sqrt{4\bar{\Omega}^{2}-\Gamma^{2}}\right]\right\} \nonumber \end{eqnarray}
 is the delay function that we look for and \begin{equation}
K(x,\tau)=\frac{2}{\sqrt{4\bar{\Omega}^{2}-\Gamma^{2}}}\phi(x,\tau)\end{equation}
 correspondingly shows the historical dependence.

If we plug the solution (\ref{eq:soln_Q}) into Eq.(\ref{eq:langevin_x}),
i.e. reduce the degree of freedom of the variable $Q(t)$ in the equation
by substituting with its formal solution, we arrive at an integro-differential
equation of only a single variable $x(t)$, which is comparable to
the feedback-containing equation of motion that appears in previous
literature \cite{ydwang07,kleckner06}. However, in the latter case
the delay function is assume to be a non-Markovian approximation \begin{equation}
h(\xi)=1-e^{-r\xi}\end{equation}
 and $K(y,\dot{y},\tau)$ is left unknown. Here, we have explicitly
implemented a self-feedback mechanism that exerts force back onto
the mass $m$ after certain delay through the use of the coupling
mass $M$. The sinusoidal factor in the delay Eq.(\ref{eq:delay})
illustrates the damped oscillating motion of the mass $M$. The implicit
time derivative of $x(t)$ within the integrand implies up to an equivalent
effect an additional friction to the motion of the mass $m$, which
damps the oscillation without increasing the target's fluctuation.
The non-integral term is Markovian and has the same order as $\bar{\omega}^{2}x$.
Though it does not appear in Eq.(\ref{eq:feedback_form}), this term
effectively reduces the oscillating frequency and shall not counteract
the feedback effect. Therefore, both terms impose noise suppression
effect to our target. The only limitation, however, is the thermal
fluctuation $F(t)$ from the coupling mass $M$ itself and it will
result in a limit for the suppression because of the competition between
this fluctuation and the effective feedback.

%%%%%%%%%%%%%%%%%%%%%%%%%%%%%%%%%%%%%%%%%%%%%%%%%

\section{\label{sec:exact_soln}Exact Solutions of the Langevin Equations}

In order to examine the validity of the above proposed mechanism and
verify the efficacy of the cooling rate, we find the analytic solutions
of the Langevin equations (\ref{eq:langevin_x}) and (\ref{eq:langevin_Q})
to reflect the displacement of the target $m$ as a response to its
own thermal noise, the motion of the ancillary mass $M$ and the thermal
noise of mass $M$. From the response function, we shall derive the
effective damping coefficient and vibrating frequency of the target
mass $m$ as well as the autocorrelation function of its coordinate.
The noise spectrum and the total noise fluctuation are then defined
upon these derived quantities.

\subsection{The Noise Spectrum}

The displacement spectrum can be written as the sum of responses of
the noise terms (the derivation is given in Appendix \ref{sec:derv_soln})
\begin{equation}
\tilde{x}(\omega)=L_{f}(\omega)\tilde{f}(\omega)+L_{F}(\omega)\tilde{F}(\omega).\label{eq:Fourier_x}\end{equation}
 Note that the two different susceptibilities \begin{equation}
L_{f}(\omega)=\frac{1}{\bar{\omega}_{e}^{2}-\omega^{2}+i\omega\gamma_{e}}\label{eq:response}\end{equation}
 and \begin{equation}
L_{F}(\omega)=\frac{g/m}{\left[\bar{\Omega}^{2}-\omega^{2}+i\omega\Gamma\right]\left[\bar{\omega}_{e}^{2}-\omega^{2}+i\omega\gamma_{e}\right]}\label{eq:response_F}\end{equation}
 reflect the system's different responses to thermal fluctuations
from the mass $m$ and the mass $M$. Here, we have defined the effective
vibrating frequency of the mass $m$ to be \begin{equation}
\bar{\omega}_{e}^{2}(\omega)=\bar{\omega}^{2}-\frac{g^{2}}{mM}\frac{\bar{\Omega}^{2}-\omega^{2}}{(\bar{\Omega}^{2}-\omega^{2})^{2}+\omega^{2}\Gamma^{2}}\label{eq:freq_eff}\end{equation}
 and the effective damping coefficient of the mass $m$ to be \begin{equation}
\gamma_{e}(\omega)=\gamma+\frac{g^{2}}{mM}\frac{\Gamma}{(\bar{\Omega}^{2}-\omega^{2})^{2}+\omega^{2}\Gamma^{2}}.\label{eq:damp_eff}\end{equation}
 Eq.(\ref{eq:Fourier_x}) means that the mechanical susceptibility
or response function of the target $m$ is adjusted because of the
dynamic coupling of the target mass $m$ to the ancillary mass $M$.
The first term represents the direct effect of the thermal bath acting
on the target mass $m$; whereas the second term represents the indirect
effect of the thermal bath onto the mass $m$ through the mediating
mass $M$ and the coupling between the two masses.

The positivity of the second term in Eq.(\ref{eq:damp_eff}) has asserted
our expectation of increasing the damping rate without increasing
thermal force. We shall also note that an additional noise source
$F(\omega)$ is imposed onto the adjusted susceptibility Eq.(\ref{eq:response})
due to the dynamic coupling. But seeing that it is divided by a frequency-squared
term, we expect it to be negligible when the target mass $m$ is not
resonating at a frequency close to that of the mass $M$.

The density noise spectrum (DNS), which is defined by the equation\begin{equation}
S_{x}(\omega)=\frac{1}{2\pi}\int_{-\infty}^{\infty}\mathrm{d}\omega^{\prime}\left\langle \tilde{x}(\omega)\tilde{x}(\omega^{\prime})\right\rangle ,\label{eq:DNS_defn}\end{equation}
 can thus be computed from Eq.(\ref{eq:Fourier_x}) \begin{equation}
S_{x}(\omega)=\frac{2k_{B}T}{m}\frac{\gamma_{e}}{(\bar{\omega}_{e}^{2}-\omega^{2})^{2}+\omega^{2}\gamma_{e}^{2}}.\label{eq:DNS}\end{equation}
 Comparing the above equation with the case when $M$ is absent, i.e.
when $\gamma_{e}$ is degenerated to $\gamma$, we observe a general
suppression at the noise peak and spreading of the noise spectrum.

\subsection{Noise Suppression}

The observable effect of the noise sources on the motion of the mass
$m$ is equivalent to the variance of the displacement of the mass
$m$ in time domain, which is defined as the average of the entire
noise spectrum, i.e. the integral of the DNS of $x$, \begin{eqnarray}
\left\langle x^{2}(t)\right\rangle  & = & \frac{1}{2\pi}\int_{-\infty}^{\infty}\mathrm{d}\omega\, S_{x}(\omega)\\
 & = & \frac{k_{B}T}{m\pi}\!\int_{-\infty}^{\infty}\!\frac{\gamma_{e}}{(\bar{\omega}_{e}^{2}-\omega^{2})^{2}+\omega^{2}\gamma_{e}^{2}}\mathrm{d}\omega.\label{eq:var_x}\end{eqnarray}
 The integral is computable after we approximate the effective damping
rate and vibrating frequency by truncating their expansions shown
in Appendix \ref{sec:derv_var}. As a result, the variance is a variable
of temperature and the spring constants \begin{equation}
\left.\left\langle x^{2}\right\rangle \right|_{k,K,g}=k_{B}T\frac{g+K}{g(k+K)+kK}.\label{eq:supp_noise}\end{equation}

The noise fluctuation depends on the three spring constants $k$,
$K$ and $g$ as its parameters, and is independent of the masses
$m$ and $M$ and the damping rate $\gamma$ and $\Gamma$ of the
resonators. We shall notice that the noise suppression effect for
the target resonator is always present for all values the spring constant
$g$ takes. This ideal result is due to the non-Markovian feedback
we derived in Sec.\ref{sec:model}, which always increases the effective
damping while retaining the same fluctuations.

Eq.(\ref{eq:supp_noise}) shows a complex relation between itself
and its three parameters of spring constants. To illustrate its behavior,
we focus on its relation with the spring constant $g$. It is a monotonic
decreasing function of $g$ and its two limiting values are \begin{eqnarray}
\left.\left\langle x^{2}\right\rangle \right\vert _{g\rightarrow0} & = & \frac{k_{B}T}{k}\\
\left.\left\langle x^{2}\right\rangle \right\vert _{g\rightarrow\infty} & = & \frac{k_{B}T}{k+K},\end{eqnarray}
 which coincides with our expectation that enlarging the constant
$g$ will render the feedback more effective due to the enlarged feedback
amplitude Eq.(\ref{eq:phi(x,t)}).

The symmetry between the target mass and the ancillary mass in the
model setup allows us to compute the variance of the ancillary mass
following the same methodology \begin{equation}
\left.\left\langle Q^{2}\right\rangle \right|_{k,K,g}=k_{B}T\frac{g+k}{g(k+K)+kK}.\end{equation}
 The limiting value of $\left\langle Q^{2}\right\rangle $ at $g\to\infty$
is identical to that of $\left\langle x^{2}\right\rangle $, which
shows that under the extremal case where the two oscillating masses
combines into one by a rigid body the limiting behaviors of the two
bodies become the same.

%%%%%%%%%%%%%%%%%%%%%%%%%%%%%%%%%%%%%%%%%%%%%%%%

\section{\label{sec:num_analysis}Numerical Analysis of Noise Suppression}

To show completely the noise suppression behavior of the two oscillating
masses, the numerical analysis is separated into two parts with each
part for each extremal end of the values of the spring constants whereas
the other parameters are set to laboratory accessible values for common
micromechanical resonators. The first case, the identical case, is
where the springs attaching the two masses to the walls share the
same Hooke's constant, i.e. $k=K$. We examine how the noise suppression
behavior is affected by varying the value of the constant $g$ of
the middle spring. The second case, the large detuning case, occurs
when the springs attaching the two masses to the walls take vastly
different values of their Hooke's constant. Again we examine the noise
suppression limit for different values taken for the constant $g$
of the middle spring.

\subsection{The Identical Case}

We assume the springs with one end fixed to walls have $k=K=1Nm^{-1}$.
The lighter target mass has $m=1\times10^{-8}kg$ and its frictional
damping rate $\gamma=0.1s^{-1}$. The heavier ancillary mass has $M=1\times10^{-6}kg$
and its frictional damping rate $\Gamma=4s^{-1}$. When free from
the stringing spring $g$, the target mass will oscillate at a natural
frequency of $10kHz$ and the ancillary mass at $1kHz$. The system's
initial temperature is set to room temperature $T=295K$.

\begin{figure}
\begin{centering}
\includegraphics[bb=30bp 30bp 308bp 241bp,width=3in]{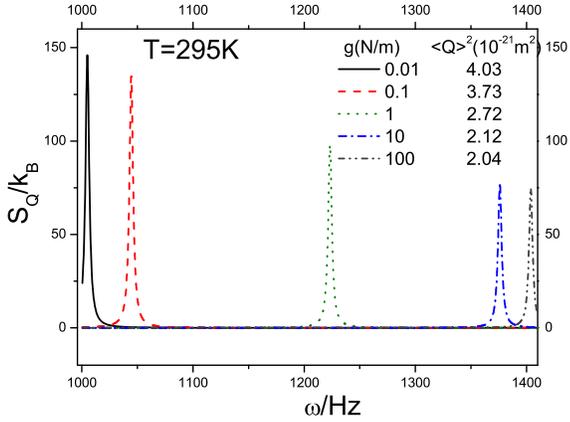} 
\par\end{centering}

\caption{(Color online) Plot of DNS of the ancillary resonator $M$ with parameters:
$k=K=1Nm^{-1}$, $M=1\times10^{-6}kg$, $m=1\times10^{-8}kg$, $\Gamma=4s^{-1}$,
$\gamma=0.1s^{-1}$, $T=295K$. Curves from top to bottom are plotted
from $g=0.01Nm^{-1}$, $0.1Nm^{-1}$, $1Nm^{-1}$, $10Nm^{-1}$, $100Nm^{-1}$.}

\label{fig:DNS_of_Q_case1} 
\end{figure}

We first look at the density noise spectrum of the ancillary resonator
when the middle spring is set to have its constant $g=0.01Nm^{-1}$,
$0.1Nm^{-1}$, $1Nm^{-1}$, $10Nm^{-1}$ and $100Nm^{-1}$ as shown
in Fig.\ref{fig:DNS_of_Q_case1}. The noise peaks at the frequencies
$\omega_{c}=1005Hz$, $1044Hz$, $1223Hz$, $1376Hz$ and $1404Hz$.

%%%%%%%%%%%%% figure of the DNS%%%%%%%%%%%%%%%%%%
%

%%%%%%%%%%%%% end of figure %%%%%%%%%%%%%%%%%%%%%

\begin{figure}
\begin{centering}
\includegraphics[bb=30bp 30bp 290bp 210bp,width=2.8in]{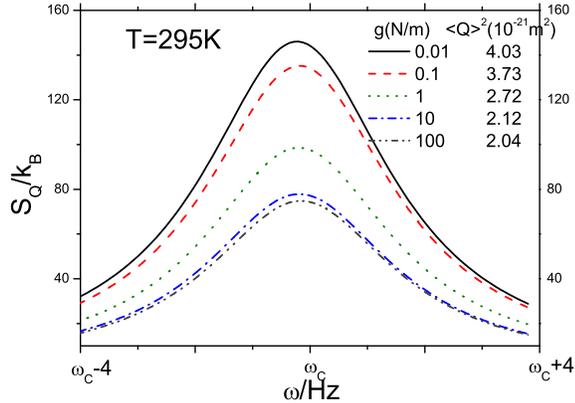} 
\par\end{centering}

\caption{(Color online) Rescaled plot of DNS of the ancillary resonator $M$
with parameters: $k=K=1Nm^{-1}$, $M=1\times10^{-6}kg$, $m=1\times10^{-8}kg$,
$\Gamma=4s^{-1}$, $\gamma=0.1s^{-1}$, $T=295K$ and peaking frequencies
shifted to the center of the plot. Curves from top to bottom are plotted
from $g=0.01Nm^{-1}$, $0.1Nm^{-1}$, $1Nm^{-1}$, $10Nm^{-1}$, $100Nm^{-1}$.}

\label{fig:DNS_of_Q_case1_shift} 
\end{figure}

We notice that when tuning the spring constant $g$, not only the
peaking frequency is shifted to the right, the peak amplitude is reduced
along with the increased value of $g$. This proves a general suppression
in noise and an equivalent cooling effect to the system. The details
of the spectrum and the spread width can be shown more apparently
when we cluster the peaks together with a common frequency $\omega_{c}$
for their corresponding peaking frequencies, which is shown in Fig.\ref{fig:DNS_of_Q_case1_shift}.
The total noise fluctuation reached after suppression can be computed
from the area under each curve in the figure, using Eq.(\ref{eq:var_x}).
The attenuated noise levels corresponding to the 5 values of the spring
$g$ are, respectively, $4.03$, $3.73$, $2.72$, $2.12$ and $2.04$
times a common factor of $10^{-21}m^{2}$.

%%%%%%%%%%%%% figure of the DNS%%%%%%%%%%%%%%%%%%%
%

%%%%%%%%%%%%% end of figure %%%%%%%%%%%%%%%%%%%%%

\begin{figure}
\begin{centering}
\includegraphics[bb=30bp 30bp 320bp 220bp,width=3in]{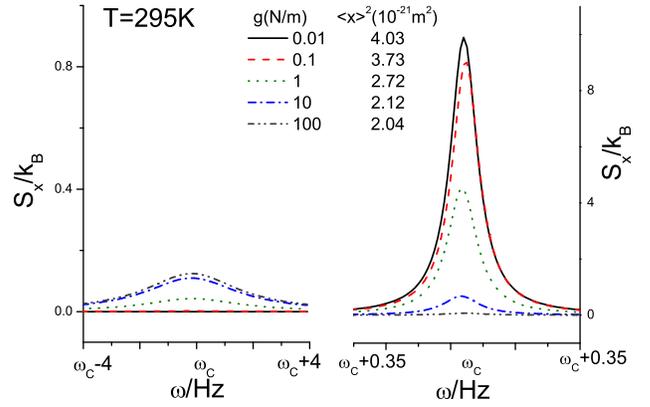} 
\par\end{centering}

\caption{(Color online) Rescaled plot of DNS of the target resonator $m$
with parameters: $k=K=1Nm^{-1}$, $M=1\times10^{-6}kg$, $m=1\times10^{-8}kg$,
$\Gamma=4s^{-1}$, $\gamma=0.1s^{-1}$, $T=295K$ and peaking frequencies
shifted to the center of the plot. The left and the right subplots
corresponds to the low and the high peak frequency groups. Curves
from top to bottom are plotted from $g=0.01Nm^{-1}$, $0.1Nm^{-1}$,
$1Nm^{-1}$, $10Nm^{-1}$, $100Nm^{-1}$.}

\label{fig:DNS_of_x_case1} 
\end{figure}

The cooling effect for the target mass $m$ is more obvious if we
examine the clustered peak plot of the target mass' density noise
spectrum shown in Fig.\ref{fig:DNS_of_x_case1}. However, differing
from the behavior of the ancillary mass shown in Fig.\ref{fig:DNS_of_Q_case1},
the target mass is resonant at two peak frequencies for each of the
5 values of the spring constant $g$. Among the pairs of peaking frequencies,
one group clusters in the low frequencies and the other spreads out
in the high frequencies.

The low frequency group, the rescaled along peak center plot shown
on the left of Fig.\ref{fig:DNS_of_x_case1}, shares exactly the same
peaking frequencies as those of the ancillary mass and we expect this
behavior takes place when the target mass is resonating with the ancillary
mass. This harmonic driven noise associates with the noise source
$\tilde{F}(\omega)$ in Eq.(\ref{eq:Fourier_x}) and, as we argue
before, does not contribute much to the overall thermal noise. The
high frequency group spreads out to peak frequencies $\omega_{c}=10.1kHz$,
$10.5kHz$, $14.2kHz$, $33.3kHz$ and $101kHz$ for the varying spring
constant $g$. The anharmonic noise with respect to the ancillary
mass associates with the noise source $\tilde{f}(\omega)$ in Eq.(\ref{eq:Fourier_x}).
Our cooling mechanism in this identical case is fairly effective,
with the noise level reached down to $4.03$, $3.73$, $2.72$, $2.12$
and $2.04$ times a common factor of $10^{-21}m^{2}$, respectively,
identical to the values of the ancillary mass.

%%%%%%%%%%%%% figure of the DNS%%%%%%%%%%%%%%%%%%%
%

%%%%%%%%%%%%% end of figure %%%%%%%%%%%%%%%%%%%%%

\subsection{The Large Detuning Case}

\begin{figure}
\begin{centering}
\includegraphics[clip,width=3in]{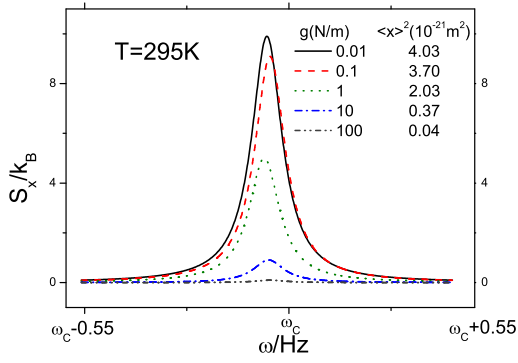} 
\par\end{centering}

\caption{(Color online) Rescaled plot of DNS of the target resonator $m$
with parameters: $K=1000Nm^{-1}$, $k=1Nm^{-1}$, $M=1\times10^{-3}kg$,
$m=1\times10^{-8}kg$, $\Gamma=4s^{-1}$, $\gamma=0.1s^{-1}$, $T=295K$
and peaking frequencies shifted to the center of the plot. Curves
from top to bottom are plotted from $g=0.01Nm^{-1}$, $0.1Nm^{-1}$,
$1Nm^{-1}$, $10Nm^{-1}$, $100Nm^{-1}$.}

\label{fig:DNS_of_x_case2} 
\end{figure}

We assume, in this case, the spring constants $K=1000Nm^{-1}$ and
$k=1Nm^{-1}$; the masses $M=1\times10^{-3}kg$ and $m=1\times10^{-8}kg$;
thus the natural oscillating frequencies for the two masses are retained.
The damping coefficients and the initial temperature are left unchanged.
Varying the spring constant of the middle spring $g$ over the same
five values gives the noise spectrum plot of the target mass $m$
shown in Fig.\ref{fig:DNS_of_x_case2}. The plot is again rescaled
to the center along the peaking frequencies $\omega_{c}=10.1kHz$,
$10.5kHz$, $14.1kHz$, $33.2kHz$ and $100.5kHz$.

We note that the noise peaks at one frequency for each value of the
spring $g$. These peaking frequencies are close to those in the identical
case above but the noise suppression rate, as we have expected, is
much better. That means the large detuning not only helps suppress
the noise source stemmed from the coupling mass $M$ to negligible
amplitude but also makes the feedback more effective for countering
the target's noise. The 5 values of the spring constant $g$ corresponds
to noise fluctuations of $4.03$, $3.70$, $2.03$, $0.37$ and $0.04$
times a common factor of $10^{-21}m^{2}$.

%%%%%%%%%%%%% figure of the DNS%%%%%%%%%%%%%%%%%%%
%

%%%%%%%%%%%%% end of figure %%%%%%%%%%%%%%%%%%%%%

\begin{figure}
\begin{centering}
\includegraphics[bb=18 25 292 229,width=3in]{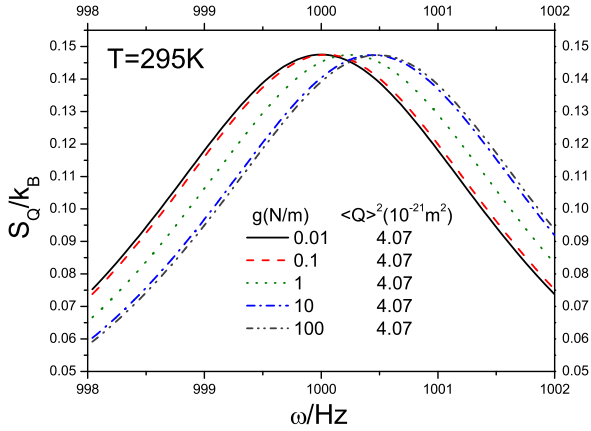} 
\par\end{centering}

\caption{(Color online) Rescaled plot of DNS of the ancillary resonator $M$
with parameters: $K=1000Nm^{-1}$, $k=1Nm^{-1}$, $M=1\times10^{-3}kg$,
$m=1\times10^{-8}kg$, $\Gamma=4s^{-1}$, $\gamma=0.1s^{-1}$, $T=295K$
and peaking frequencies shifted to the center of the plot. Curves
from left to right are plotted from $g=0.01Nm^{-1}$, $0.1Nm^{-1}$,
$1Nm^{-1}$, $10Nm^{-1}$, $100Nm^{-1}$.}

\label{fig:DNS_of_Q_case2} 
\end{figure}

Fig.\ref{fig:DNS_of_Q_case2} shows the plot of the ancillary resonator's
noise spectrum, again rescaled to the center along the peaking frequencies.
These peaking frequencies are very close to those of the target mass.
We predict that the large detuning between the springs $K$ and $k$
puts the ancillary mass $M$ into a particularly passive role that
reflects the minute motion of the target. This also helps explain
why the noise suppression is especially effective in this case by
the fact that the mass $M$ has its speed comparable to the target
$m$ but in an opposite direction such that the force it exerts through
the spring $g$ can favorably counteract the motion of the target
$m$. Nonetheless, the reinforced role that the ancillary mass $M$
plays means that itself does not belong to the target system. As shown
in the figure, the ancillary mass almost retains its original noise
level throughout the varying values of the spring $g$.

%%%%%%%%%%%%% figure of the DNS%%%%%%%%%%%%%%%%%%

%

%%%%%%%%%%%%% end of figure %%%%%%%%%%%%%%%%%%%%%

\section{\label{sec:momentum-noise}Noise in Momentum Space}

\subsection{The Model with Harmonic Driving Force}

The effective temperature is usually defined according to the equipartition
theorem through the equality between the thermal energy and the mechanical
potential energy. The intrinsic nature of our model renders the stored
potential energy of the target and of the environment in their shared
link, the middle spring $g$, inseparable and thus forbids our discussion
of the effective temperature in the coordinate space. To make our
model relevant to the discussion of cooling, we examine the system
dynamics in its momentum space and define the effective temperature
according to the kinetic energy, i.e. the momentum fluctuation, of
the target.

In addition, to follow the convention along previous literature, a
harmonic driving force term \begin{equation}
G(t)=F_{0}e^{-i\omega_{0}t},\end{equation}
 where $\omega_{0}$ is the driving frequency, should be added to
the feedback loop (Cf. Fig.\ref{fig:model}). In our case, this harmonic
force appears as an extra term in the equation of motion of the ancillary
resonator $M$.

Following the above arguments, we rewrite the coupled system of classical
Langevin equations Eq.(\ref{eq:langevin_x}) and Eq.(\ref{eq:langevin_Q})
as\begin{eqnarray}
\dot{p}+\gamma p+m\bar{\omega}^{2}x-gQ & = & mf\label{eq:Langevin_mo_x}\\
\dot{P}+\Gamma P+M\bar{\Omega}^{2}Q-gx & = & M(F+G(t))\label{eq:Langevin_mo_Q}\end{eqnarray}
 where $p=m\dot{x}$ and $P=M\dot{Q}$ denote, respectively, the momenta
of the target resonator and the ancillary resonator. The thermal fluctuations
$f$ and $F$ still obey the same set of relations Eq.(\ref{eq:FD-relation})
and Eq.(\ref{eq:FD-relation_F}).

Consequently, we have a modified linear response (Cf. Eq.(\ref{eq:Fourier_x}),
Eq.(\ref{eq:response}) and Eq.(\ref{eq:response_F})) after combining
the system of equations in the frequency domain (see Appendix \ref{sec:derv_eff_temp})\begin{multline}
\tilde{p}(\omega)=i\omega m\Bigl[L_{f}(\omega)\tilde{f}(\omega)+L_{F}(\omega)[\tilde{F}(\omega)\\
+2\pi F_{0}\delta(\omega+\omega_{0})]\Bigr]\label{eq:Fourier_p}\end{multline}
 where $L_{f}(\omega)$ and $L_{F}(\omega)$ are the susceptibilities
Eq.(\ref{eq:response}) and Eq.(\ref{eq:response_F}). We can thus
arrive at the noise spectrum in momentum, after following the same
routine of computations,\begin{eqnarray}
S_{p}(\omega) & = & \frac{1}{2\pi}\int_{-\infty}^{\infty}\mathrm{d}\omega'\left\langle \tilde{p}(\omega)\tilde{p}(\omega')\right\rangle \\
 & = & 2k_{B}Tm\frac{\omega^{2}\gamma_{e}}{(\bar{\omega}_{e}^{2}-\omega^{2})^{2}+\omega^{2}\gamma_{e}^{2}}+\nonumber \\
 &  & 2\pi F_{0}^{2}\omega_{0}^{2}m^{2}\delta(\omega+\omega_{0})L_{F}(\omega_{0})L_{F}(-\omega_{0})\end{eqnarray}

\subsection{The Effective Temperature }

The square of momentum of the target is defined similarly as in Eq.(\ref{eq:var_x})\begin{equation}
\left\langle p^{2}\right\rangle =\frac{1}{2\pi}\int_{-\infty}^{\infty}\mathrm{d}\omega\, S_{p}(\omega).\end{equation}
 which can be separated into two parts: the variance of momentum,
i.e. the fluctuation or noise in momentum space,\begin{equation}
\left\langle \left(\Delta p\right)^{2}\right\rangle =\frac{k_{B}T}{\pi}m\int_{-\infty}^{\infty}\mathrm{d}\omega\frac{\omega^{2}\gamma_{e}}{(\bar{\omega}_{e}^{2}-\omega^{2})^{2}+\omega^{2}\gamma_{e}^{2}}\label{eq:DNS_mo}\end{equation}
and the mean squared, i.e., the square of the steady state value of
momentum, \begin{eqnarray}
\left\langle p\right\rangle ^{2} & = & F_{0}^{2}\omega_{0}^{2}m^{2}L_{F}(\omega_{0})L_{F}(-\omega_{0}).\end{eqnarray}
Note that the harmonic force term only contributes to the first moment
of the momentum.

Since thermal dissipation only induces variance of the target's momentum,
the effective temperature of the target can inversely be determine
by the variance of the kinetic energy through the energy equipartition
theorem \begin{equation}
T_{\mathrm{eff}}=\frac{\left\langle \left(\Delta p\right)^{2}\right\rangle }{2k_{B}m}.\end{equation}
Eq.(\ref{eq:DNS_mo}) is independent of the two parameters $F_{0}$
and $\omega_{0}$ of the harmonic driving force. We conclude that
the harmonic force bears no effect in reducing the system's effective
temperature and hence in cooling. Eq.(\ref{eq:DNS_mo}) is also not
analytically integrable, numerical integration shows the integral
takes value similar to the case when the spring constant $g=0$. That
means the feedback is only effective in the displacement domain but
not in the momentum domain.

\section{Discussion and remarks}

In summary, we have proposed a theoretical model to demonstrate the
general self-feedback type noise suppression technique through the
coupling between the target resonator and an adjuvant system. In particular,
such a self-feedback is achieved through the dynamics of the adjuvant
system. The explicit delay function has been given and its efficacy
in reducing the noise spectrum verified. We have also used numeral
results to confirm our observations.

Before concluding this paper, we add some remarks as follows:

First, the theory of thermal equilibrium state can be used to explain
why the system comprising two resonating masses has its noise reduced to
a limit when the spring constant $g$ tends from zero to infinity.
If $g$ were zero, the stringing spring between the two masses would
disappear and the system would comprise two independent resonators,
each of which will attain its thermal equilibrium over time. In this
case, the system has two degrees of freedom. According to the energy
equipartition theorem, the system must contain twice as much energy
as $k_{B}T/2$. For the other limit, when $g$ tends to infinity,
the two resonators become a rigid body and possess only one degree
of freedom, which means the system would contain only $k_{B}T/2$
of energy. This explicit picture clarifies how the combined system
reduces its energy over the increasing value of $g$.

However, how each resonator in the combined system reduces its own
energy is an open question though we can intuitively think the heat
bath may play crucial role in the asymmetric thermalization of the
two resonators as tow open system. Another remark is that the energy
equipartition theorem is applicable only after the system enters thermal
equilibrium state. When interaction occurs between the two degrees
of freedom contributed by the two masses, such an illustration is
not appropriate. This is the reason why the nonlinear character of
the noise suppression rate cannot be explained by the energy equipartition
theorem alone.

In addition, we have assumed there does not exist correlation between
the two noise sources for the two masses, i.e. $\left\langle f(t)F(t^{\prime})\right\rangle =0$.
However, when the two masses oscillate very closely with each other,
the above assumption based on independent thermal environments will
not hold and we need to consider the case where \begin{equation}
\left\langle f(t)F(t^{\prime})\right\rangle \neq0.\end{equation}

Another setting we shall consider is when the motions of the two masses
are quantized: we need to change the fluctuation-dissipation relations
Eq.(\ref{eq:FD-relation}) and Eq.(\ref{eq:FD-relation_F}) to \begin{eqnarray}
\left\langle f(t)f(t^{\prime})\right\rangle  & \propto & W(t,t^{\prime},\hbar),\\
\left\langle F(t)F(t^{\prime})\right\rangle  & \propto & U(t,t^{\prime},\hbar).\end{eqnarray}
 where $W$ and $U$ are not simply $\delta$-functions and depends
on the Planck constant $\hbar$. We will present the general investigation
on the case with quantum fluctuations and thermal bath correlations,
but here we have concentrated on the simple case.

%%%%%%%%%%%%%%%%%%%%%%%%%%%%%%%%%%%%%%%%%%%%%%%%%

\begin{acknowledgments}
The authors thank Yong Li of University of Basel and Ying Dan Wang
of NTT Basic Research Laboratories for helpful discussions. This work
is supported by the NSFC with Grants No.90203018, No.10474104, and
No.60433050. It is also funded by the National Fundamental Research
Program of China with Grants No.2001CB309310 and No.2005CB724508. 
\end{acknowledgments}
%%%%%%%%%%%%%%%%%%%%%%%%%%%%%%%%%%%%%%%%%%%%%%%%%

\appendix
%dummy comment inserted by tex2lyx to ensure that this paragraph is not empty

\section{\label{sec:derv_delay}Derivation of the Feedback Delay Function}

Fourier transforming Eq.(\ref{eq:langevin_Q}) gives \begin{equation}
\tilde{Q}(\omega)=\frac{g\tilde{x}(\omega)/M+\tilde{F}(\omega)}{\bar{\Omega}^{2}-\omega^{2}+i\omega\Gamma}\label{eq:Q_tfm}\end{equation}
 where we use $\tilde{x}$, $\tilde{Q}$ and $\tilde{F}$ to denote
the Fourier transforms of $x$, $Q$ and $F$, respectively, \begin{eqnarray}
\tilde{x}(\omega) & = & \int_{-\infty}^{\infty}\mathrm{d}t\, x(t)e^{-i\omega t},\\
\tilde{Q}(\omega) & = & \int_{-\infty}^{\infty}\mathrm{d}t\, Q(t)e^{-i\omega t},\\
\tilde{F}(\omega) & = & \int_{-\infty}^{\infty}\mathrm{d}t\, F(t)e^{-i\omega t}.\end{eqnarray}
 Substituting the first two transform of the above into Eq.(\ref{eq:Q_tfm})
and taking the inverse Fourier transform of $\tilde{Q}$, we get \begin{equation}
Q(t)=\frac{1}{2\pi}\int_{-\infty}^{\infty}\mathrm{d}\tau\mathrm{d}\omega\frac{gx(\tau)/M+F(\tau)}{(\bar{\Omega}^{2}-\omega^{2})+i\Gamma\omega}e^{-i\omega(\tau-t)}.\end{equation}
 The linear integral with respect to $\omega$ can be viewed as a
contour integral along a semicircle in the upper half-plane with two
poles at $\omega_{\pm}=[i\Gamma\pm\sqrt{4\bar{\Omega}^{2}-\Gamma^{2}}]/2$.
Assuming $\Gamma\ll\bar{\Omega}$, the corresponding residues are,
\begin{eqnarray}
R_{\pm}(t) & = & \mp\frac{1}{\sqrt{4\bar{\Omega}^{2}-\Gamma^{2}}}\theta(t-\tau)\times\\
 &  & \exp\left[-\frac{1}{2}\left(\Gamma\mp i\sqrt{4\bar{\Omega}^{2}-\Gamma^{2}}\right)(t-\tau)\right]\nonumber \end{eqnarray}
 where $\theta(t-\tau)$ denotes the unit step function originated
from the positive locus of the integration path. Using Cauchy's theorem,
$Q(t)$ is reduced to a single-integral form \begin{eqnarray}
Q(t) & = & \frac{2}{\sqrt{4\bar{\Omega}^{2}-\Gamma^{2}}}\int_{-\infty}^{t}\mathrm{d}\tau\,\phi(x,\tau)\times\\
 &  & \exp\left[-\frac{1}{2}\Gamma(t-\tau)\right]\sin\left[\frac{1}{2}\sqrt{4\bar{\Omega}^{2}-\Gamma^{2}}(t-\tau)\right]\nonumber \end{eqnarray}
 where we have used the shorthand $\phi(x,t)$ as defined in Eq.(\ref{eq:phi(x,t)}).
To write $Q(t)$ in our desired form, we further integrate by parts
with respect to $(t-\tau)$ \begin{eqnarray}
Q(t) & = & \frac{-2}{\sqrt{4\bar{\Omega}^{2}-\Gamma^{2}}}\int_{0}^{\infty}\mathrm{d}(t-\tau)\phi(x,t)\\
 &  & \times\exp\left[-\frac{1}{2}\Gamma(t-\tau)\right]\sin\left[\frac{t-\tau}{2}\sqrt{4\bar{\Omega}^{2}-\Gamma^{2}}\right]\nonumber \\
 & = & \frac{1}{\bar{\Omega}^{2}}\phi(x,t)+\frac{2}{\sqrt{4\bar{\Omega}^{2}-\Gamma^{2}}}\times\\
 &  & \int_{0}^{\infty}\mathrm{d}(t-\tau)\frac{\mathrm{d}\phi(x,\tau)}{\mathrm{d}(t-\tau)}2\exp\left[-\frac{1}{2}\Gamma(t-\tau)\right]\times\nonumber \\
 &  & \left\{ \frac{\exp\left[\frac{1}{2}i\sqrt{4\bar{\Omega}^{2}-\Gamma^{2}}(t-\tau)\right]}{\Gamma-i\sqrt{4\bar{\Omega}^{2}-\Gamma^{2}}}\right.+\nonumber \\
 &  & \left.\frac{\exp\left[-\frac{1}{2}i\sqrt{4\bar{\Omega}^{2}-\Gamma^{2}}(t-\tau)\right]}{\Gamma+i\sqrt{4\bar{\Omega}^{2}-\Gamma^{2}}}\right\} \nonumber \end{eqnarray}
 where the factor in the last two lines constitute the function $h(t-\tau)$
as defined in Eq.(\ref{eq:delay}).

%%%%%%%%%%%%%%%%%%%%%%%%%%%%%%%%%%%%%%%%%%%%%%%%%

\section{\label{sec:derv_soln}Derivation of the Response Function and the
Noise Spectrum}

To find the response function, we first take the Fourier transform
of Eq.(\ref{eq:langevin_Q}) and Eq.(\ref{eq:langevin_x}) to get,
respectively, Eq.(\ref{eq:Q_tfm}) and \begin{equation}
\tilde{x}(\omega)=\frac{g\tilde{Q}(\omega)/m+\tilde{f}(\omega)}{\bar{\omega}^{2}-\omega^{2}+i\omega\gamma}\label{eq:x_tfm}\end{equation}
 where we let $\tilde{x}$, $\tilde{Q}$ and $\tilde{f}$ denote the
Fourier transforms of $x$, $Q$ and $f$ as we did in Appendix \ref{sec:derv_delay}.
Substituting Eq.(\ref{eq:Q_tfm}) into Eq.(\ref{eq:x_tfm}), we have\begin{equation}
\tilde{x}(\omega)=\left(\frac{g}{m}\frac{g\tilde{x}(\omega)/M+\tilde{F}(\omega)}{\bar{\Omega}^{2}-\omega^{2}+i\omega\Gamma}+\tilde{f}(\omega)\right)(\bar{\omega}^{2}-\omega^{2}+i\omega\gamma)^{-1}\end{equation}
Reshuffling the terms and putting again $\tilde{x}(\omega)$ on one
hand side, the equation becomes \begin{equation}
\tilde{x}(\omega)=\frac{\tilde{f}(\omega)+g\tilde{F}(\omega)/[m(\bar{\Omega}^{2}-\omega^{2}+i\omega\Gamma)]}{\bar{\omega}^{2}-\omega^{2}+i\omega\gamma-g^{2}/[mM(\bar{\Omega}^{2}-\omega^{2}+i\omega\Gamma)]}.\end{equation}
 To find explicitly the response function, we group together the real
terms and the imaginary terms in the denominator to write $\tilde{x}(\omega)$
in a familiar form similar to the case where the coupling mass $M$
were not present; whence the denominator becomes\begin{multline}
\left[\bar{\omega}^{2}-\frac{g^{2}}{mM}\frac{\bar{\Omega}^{2}-\omega^{2}}{(\bar{\Omega}^{2}-\omega^{2})^{2}+\omega^{2}\Gamma^{2}}\right]-\omega^{2}\\
+i\omega\left[\gamma+\frac{g^{2}}{mM}\frac{\Gamma}{(\bar{\Omega}^{2}-\omega^{2})^{2}+\omega^{2}\Gamma^{2}}\right]\end{multline}
 and we can define the terms in the two brackets as in Eq.(\ref{eq:freq_eff})
and Eq.(\ref{eq:damp_eff}).

Using Eq.(\ref{eq:Fourier_x}), the associated fluctuation-dissipation
relations of Eq.(\ref{eq:FD-relation}) and Eq.(\ref{eq:FD-relation_F})
in the frequency domain \begin{eqnarray}
\left\langle \tilde{f}(\omega)\tilde{f}(\omega^{\prime})\right\rangle  & = & 4\pi k_{B}T\frac{\gamma}{m}\delta(\omega+\omega^{\prime}),\label{eq:freq-FD}\\
\left\langle \tilde{F}(\omega)\tilde{F}(\omega^{\prime})\right\rangle  & = & 4\pi k_{B}T\frac{\Gamma}{M}\delta(\omega+\omega^{\prime}),\label{eq:freq-FD_F}\end{eqnarray}
 and the independence between the two noise sources, we find by using
definition Eq.(\ref{eq:DNS_defn}) \begin{equation}
S_{x}(\omega)=2k_{B}T\left[\frac{\gamma}{m}L_{f}(\omega)L_{f}(-\omega)+\frac{\Gamma}{M}L_{F}(\omega)L_{F}(-\omega)\right]\label{eq:DNS_LL}\end{equation}
 where $L_{f}(\omega)$ and $L_{F}(\omega)$ are defined in Eq.(\ref{eq:response})
and Eq.(\ref{eq:response_F}). Seeing that \begin{equation}
L_{F}(\omega)=\frac{g/m}{\bar{\Omega}^{2}-\omega^{2}+i\omega\Gamma}L_{f}(\omega),\end{equation}
 we can factor out $L_{f}(\omega)L_{f}(-\omega)/m$ from the bracket
in Eq.(\ref{eq:DNS_LL}) \begin{eqnarray}
S_{x}(\omega) & = & \frac{2k_{B}T}{m}L_{f}(\omega)L_{f}(-\omega)\times\\
 &  & \left[\gamma+\frac{g^{2}\Gamma/mM}{\bar{\Omega}^{2}-\omega^{2}+i\omega\Gamma}\frac{1}{\bar{\Omega}^{2}-\omega^{2}-i\omega\Gamma}\right]\nonumber \\
 & = & \frac{2k_{B}T}{m}\left[\frac{1}{(\bar{\omega}_{e}^{2}-\omega^{2})^{2}+\omega^{2}\gamma_{e}^{2}}\right]\times\\
 &  & \left[\gamma+\frac{g^{2}}{mM}\frac{\Gamma}{(\bar{\Omega}^{2}-\omega^{2})^{2}+\omega^{2}\Gamma^{2}}\right]\nonumber \end{eqnarray}
 which equals to Eq.(\ref{eq:DNS}).

\section{\label{sec:derv_var}Derivation of the Variance of Displacement}

The integrand of Eq.(\ref{eq:var_x}) has highest order of $\omega^{4}$
in the denominator and can not be integrated analytically. To make
the integrand integrable, we introduce a parameter $\mu(K)=g/(g+K)$
as a variable of the spring constant $K$ and approximate the integrand
by expanding $\bar{\omega}_{e}^{2}$ with respect to $\mu$ up to
2nd order in the vicinity of $\mu=0$, in other words, when $K\rightarrow\infty$.
That is, we let \begin{equation}
\bar{\omega}_{e}^{2}(\mu)\approx\left.\bar{\omega}_{e}^{2}\right\vert _{\mu=0}+\left.\frac{\mathrm{d}\bar{\omega}_{e}^{2}}{\mathrm{d}\mu}\right\vert _{\mu=0}\!\!\mu+\frac{1}{2}\left.\frac{\mathrm{d}^{2}\bar{\omega}_{e}^{2}}{\mathrm{d}\mu^{2}}\right\vert _{\mu=0}\!\!\mu^{2}.\end{equation}
 To calculate the three terms in the expansion, we first substitute
$K$ with $g(1-\mu)/\mu$ in Eq.(\ref{eq:freq_eff}), \begin{equation}
\bar{\omega}_{e}^{2}(\mu)=\bar{\omega}^{2}-\frac{g^{2}}{mM}\frac{gM^{-1}\mu^{-1}-\omega^{2}}{(gM^{-1}\mu^{-1}-\omega^{2})^{2}+\omega^{2}\Gamma^{2}}.\label{eq:exact_omega}\end{equation}
 Since the lowest-order term in the numerator and the denominator
is, respectively, $\mu^{-1}$ and $\mu^{-2}$, the second term of
the above formula goes to 0 when $\mu\rightarrow0$. Thus we find
the first term in the expansion $\left.\bar{\omega}_{e}^{2}\right\vert _{\mu=0}=\bar{\omega}^{2}$.
The first-order derivative of $\bar{\omega}_{e}^{2}(\mu)$ reads \begin{equation}
\frac{\mathrm{d}\bar{\omega}_{e}^{2}}{\mathrm{d}\mu}=\frac{g^{3}}{mM^{2}}\frac{\mu^{-2}\left[-(gM^{-1}\mu^{-1}-\omega^{2})^{2}+\omega^{2}\Gamma^{2}\right]}{\left[(gM^{-1}\mu^{-1}-\omega^{2})^{2}+\omega^{2}\Gamma^{2}\right]^{2}}.\end{equation}
 Again by comparing the coefficients of the lowest-order terms in
the numerator and the denominator, we shall see the second term in
the expansion $\mathrm{d}\bar{\omega}_{e}^{2}/\mathrm{d}\mu|_{\mu=0}=-g/m$.
The third term in the expansion can be obtained similarly by observing
the limiting behavior of the numerator and the denominator of the
second-order derivative of $\bar{\omega}_{e}^{2}(\mu)$ \begin{equation}
\left.\frac{\mathrm{d}^{2}\bar{\omega}_{e}^{2}}{\mathrm{d}\mu^{2}}\right\vert _{\mu=0}=\lim_{\mu\rightarrow0}\frac{g^{3}}{mM^{2}}\frac{-2gM^{-1}\mu^{-4}(gM^{-1}\mu^{-1}-\omega^{2})^{2}}{\left[(gM^{-1}\mu^{-1}-\omega^{2})^{2}+\omega^{2}\Gamma^{2}\right]^{3}},\end{equation}
 which reads $-2(M/m)\omega^{2}$ after taking the limit. Combining
the results, we have \begin{equation}
\bar{\omega}_{e}^{2}\approx\bar{\omega}^{2}-\frac{g}{m}\mu-\frac{M}{m}\omega^{2}\mu^{2}.\label{eq:appr_omega}\end{equation}

The truncation error rate introduced in this approximation, by comparing
Eq.( \ref{eq:exact_omega}) and Eq.(\ref{eq:appr_omega}), is \begin{equation}
\mathrm{erf}=\frac{\Delta\bar{\omega}_{e}^{2}}{\bar{\omega}_{e}^{2}}=\frac{\bar{\omega}_{e}^{2}-\left[\bar{\omega}^{2}-(g\mu-M\omega^{2}\mu^{2})/m\right]}{\bar{\omega}_{e}^{2}}.\end{equation}
 Expanding $\bar{\omega}_{e}^{2}$ and $\mu$ gives \begin{eqnarray}
\mathrm{erf} & = & \left\{ \left[\frac{g^{2}}{g+K}+\frac{M\omega^{2}g^{2}}{(g+K)^{2}}\right](F^{2}+M^{2}\omega^{2}\Gamma^{2})-g^{2}F\right\} \nonumber \\
 &  & \times\left[(g+k)(F^{2}+M^{2}\omega^{2}\Gamma^{2})-g^{2}F\right]^{-1}\end{eqnarray}
 where we have used the shorthand $F=g+K-M\omega^{2}$. Since we only
consider the usual cases with $\Gamma\ll\omega$, the terms containing
the damping coefficient can be omitted\begin{equation}
\mathrm{erf}=\frac{\left[g^{2}/(g+K)+M\omega^{2}g^{2}/(g+K)^{2}\right]F-g^{2}}{(g+k)F-g^{2}}.\end{equation}
 Expanding $F$ and multiplying the numerator and denominator by $(g+K)^{2}$,
we find \begin{equation}
\left|\frac{\Delta\bar{\omega}_{e}^{2}}{\bar{\omega}_{e}^{2}}\right|=\frac{g^{2}M^{2}\omega^{4}}{(g+K)^{2}\left[g(k+K-M\omega^{2})+k(K-M\omega^{2})\right]}\end{equation}
 and at the other extreme of the expansion $\mu=1$ whence $K=0$
and $g$ can take any value \begin{equation}
\mathrm{erf}|_{K=0}=\frac{M^{2}\omega^{4}}{g(k-M\omega^{2})-kM\omega^{2}}.\end{equation}
 Therefore, the error can be sufficiently suppressed if we let $g\to\infty$
and the expansion of $\bar{\omega}_{e}^{2}$around $\mu=1$ is validated.

Following the same reasoning and procedure, we can approximate the
effective damping coefficient Eq.(\ref{eq:damp_eff}), \begin{equation}
\gamma_{e}\approx\gamma+\frac{M\Gamma}{m}\mu^{2}\label{eq:appr_gamma}\end{equation}
 which becomes $\omega$-independent. Substituting Eq.(\ref{eq:appr_omega})
and Eq.(\ref{eq:appr_gamma}) into Eq.(\ref{eq:var_x}), we get, after
minor algebra, \begin{equation}
\left\langle x^{2}(t)\right\rangle =\frac{k_{B}T\gamma^{\prime2}}{m\pi\gamma_{e}}\!\int_{-\infty}^{\infty}\!\frac{1}{(\bar{\omega}^{\prime2}-\omega^{2})^{2}+\omega^{2}\gamma^{\prime2}}\mathrm{d}\omega\label{eq:appr_var}\end{equation}
 where \begin{eqnarray}
\gamma^{\prime} & = & \frac{\gamma+(M\Gamma/m)\mu^{2}}{(M/m)\mu^{2}+1}\\
\bar{\omega}^{\prime} & = & \sqrt{\frac{\bar{\omega}^{2}-(g/m)\mu}{(M/m)\mu^{2}+1}}\end{eqnarray}
 can be considered the effective damping coefficient and vibrating
frequency of the mass $m$ after the approximation. We can apply a
limiting process to Eq.(\ref{eq:appr_var}) \begin{equation}
\left\langle x^{2}(t)\right\rangle =\frac{k_{B}T\gamma^{\prime2}}{m\pi\gamma_{e}}\lim_{t\to0}\int_{-\infty}^{\infty}\frac{e^{-i\omega t}}{(\bar{\omega}^{\prime2}-\omega^{2})^{2}+\omega^{2}\gamma^{\prime2}}\mathrm{d}\omega\end{equation}
 and compute the integral using the theorem of residues as we did
in Appendix \ref{sec:derv_delay} \begin{eqnarray}
\left\langle x^{2}(t)\right\rangle  & = & \frac{k_{B}T\gamma^{\prime}}{m\gamma_{e}\bar{\omega}^{\prime2}}\\
 & = & k_{B}T\frac{g+K}{g(k+K)+kK}.\end{eqnarray}

\section{\label{sec:derv_eff_temp}Derivation of Effective Temperature}

Fourier transforming the Langevin equations in the momentum space
Eq.(\ref{eq:Langevin_mo_x}) and Eq.(\ref{eq:Langevin_mo_Q}) and
recognizing \begin{eqnarray}
\tilde{p}(\omega) & = & i\omega\tilde{x}(\omega)\\
\tilde{P}(\omega) & = & i\omega\tilde{Q}(\omega),\end{eqnarray}
 we have\begin{align}
(\bar{\omega}^{2}-\omega^{2}+i\omega\gamma)\tilde{p}(\omega)-\frac{g}{M}\tilde{P}(\omega) & =i\omega m\tilde{f}(\omega)\\
(\bar{\Omega}^{2}-\omega^{2}+i\omega\Gamma)\tilde{P}(\omega)-\frac{g}{m}\tilde{p}(\omega) & =i\omega M(\tilde{F}(\omega)\nonumber \\
 & +2\pi F_{0}\delta(\omega+\omega_{0})).\end{align}

We can follow the lines in Appendix \ref{sec:derv_soln} at this point
to get the Fourier transform of the momentum of the target resonator\begin{multline}
\tilde{p}(\omega)=i\omega\left[m\tilde{f}(\omega)+g\frac{\tilde{F}(\omega)+2\pi F_{0}\delta(\omega+\omega_{0})}{\bar{\Omega}-\omega^{2}+i\omega\Gamma}\right]\\
\times\left[\bar{\omega}^{2}-\omega^{2}+i\omega\gamma-\frac{g}{mM}\frac{1}{\bar{\Omega}-\omega^{2}+i\omega\Gamma}\right]^{-1}\end{multline}
 which can be written in a compact form as in Eq.(\ref{eq:Fourier_p})
after grouping the terms. The density noise spectrum in momentum can
then be computed\begin{align}
S_{p} & (\omega)=\frac{1}{2\pi}\int_{-\infty}^{\infty}\mathrm{d}\omega'\left\langle \tilde{p}(\omega)\tilde{p}(\omega')\right\rangle \\
 & =2k_{B}T\omega^{2}m^{2}\left[\frac{\gamma}{m}L_{f}(\omega)L_{f}(-\omega)+\frac{\Gamma}{M}L_{F}(\omega)L_{F}(-\omega)\right]\\
 & +2\pi F_{0}^{2}\omega^{2}m^{2}\delta(\omega+\omega_{0})L_{F}(\omega)L_{F}(-\omega_{0})\nonumber \\
 & =2k_{B}Tm\frac{\omega^{2}\gamma_{e}}{(\bar{\omega}_{e}^{2}-\omega^{2})^{2}+\omega^{2}\gamma_{e}^{2}}+\\
 & +2\pi F_{0}^{2}\omega_{0}^{2}m^{2}\delta(\omega+\omega_{0})L_{F}(\omega_{0})L_{F}(-\omega_{0})\nonumber \end{align}

\end{document}